\documentstyle[11pt,newpasp,twoside,epsf]{article}
\def\edcomment#1{\iffalse\marginpar{\raggedright\sl#1\/}\else\r
elax\fi}
\reversemarginpar

\begin{document}
\title{Substellar Companions to Nearby Stars from NICMOS Surveys}
\author{Glenn Schneider}
\affil{Steward Obs., U. Arizona, 933 N. Cherry Ave., Tucson, AZ 
85721 
USA}
\author{Eric E. Becklin \& Patrick J. Lowrance}
\affil{UCLA, 405 Hilgard Ave., Los Angeles, CA 90095 USA}
\author{Bradford A. Smith}
\affil{U. Hawaii, 82-6012 Puuhonua Rd. Naoopoo, HI 96704 USA}
\author{and the NICMOS IDT/EONS and HST GO/8176 Teams}

\begin{abstract}
A coronagraphic imaging survey of 65 nearby stars was conducted 
using the Near Infrared Camera and Multi-Object Spectrometer (NICMOS) 
on the Hubble Space Telescope by the Instrument Definition Team.  Using 
these guaranteed time observations we searched for  very low 
mass stellar and substellar (extra-solar giant planet and brown dwarf) 
companions to young main sequence and late M dwarf stars.  
Nineteen additional stars, similarly observed in our circumstellar disk 
program, were also examined for evidence of companions.  
We discuss the large depth and dynamic range of these surveys which
have given rise to several putative and confirmed detections.  
Using follow-up spectrographic observations obtained with
the Space Telescope Imaging Spectrograph (STIS) we have begun to 
characterize the physical nature of the objects found in the companion search program. 
\end{abstract}

\section{Introduction}
Is there a continuity of companion objects across the sub-stellar 
mass-spectrum bridging the stellar main sequence  into the 
planetary domain? 
If so, in what sort of local environments will these objects form? At 
what distances will  they be found from their primaries and how might 
this be affected by the characteristics of the primary and  companion 
objects and of the circumstellar regions?  What implications will the 
discovery and characterization of  such objects have for our 
understanding of the formation mechanisms of extra-solar planetary
systems? 

To begin answering these fundamental questions we have searched
for candidate brown dwarf and extra-solar giant planet (EGP)
companions and circumstellar debris disks in a direct imaging
survey using the HST/NICMOS coronagraph. Our survey of 84 stars is
sensitive to transitional and substellar companions over a broad 
range of masses and separations.   Our disk imaging program is 
discussed elsewhere in these conference 
proceedings (Weinberger et al. 2000); here we concentrate on our 
companion survey. With the NICMOS coronagraph, and exploiting the highly 
stable characteristics of the HST+NICMOS point spread function (PSF)
on sub-orbit timescales, 
we are able to image unresolved objects 13 to 15 magnitudes fainter than
the occulted stars at distances of only a few arc seconds 
(Schneider 1998), thus identifying companion candidates for 
follow-up observations.  To unequivocally establish the physical
association of a companion candidate with its suspected primary,
common proper motions must be demonstrated through astrometric 
measurements. Spectroscopy of a putative companion is 
necessary to ascertain its physical nature.

\section{The NICMOS Coronagraphic Survey}
The first phase of this program, designed to identify candidate 
companion objects, was carried out as a single color survey 
using the F160W filter (a very close analog to H-band) where 
the instrumental performance for low-mass  point-source detection 
is optimized (Lowrance et al 1998). Each star was imaged coronagraphically 
at two field orientations (spacecraft roll angles) in the 
same orbit. The coronagraph reduces the background 
scattered light from the primary target by factors of several 
to more than an order of magnitude, varying with  radial distance.  
The  PSF of the primary and residual 
scattered light artifacts remain fixed as the  detector corotates with 
the spacecraft. Subtraction of the two differentially rolled frames 
further reduces the background light by factors of several hundred in a 
spatially dependent fashion. This leaves ``conjugate'' (positive and  
negative) images of any astronomical sources rolled about the target 
axis which may be later recombined. By combining all frames 
in the one orbit per target allocated for the NICMOS coronagraphic 
survey, companion objects as faint as H $\sim$ 23 can be detected a 
few arc seconds from many of our primary targets (Schneider et al 1998). 
For the youngest stars in our census, evolutionary (cooling) models 
by Burrows et al (1997) and others indicate that this sensitivity limit 
corresponds to companion masses extending into the Jovian domain. 
This differential imaging strategy was generally adopted in 
our coordinated observing programs under the umbrella of 
the NICMOS IDT's Environments of Nearby Stars investigations.
While different in detail, those programs may be 
be summarized as follows: 

A Search For Massive Jupiters. Thirty-eight very young 
main-sequence stars with a mean distance of $\sim$ 30pc were observed.  
The median age for the selected targets with well-established ages is 
$\sim$  90 Myrs, with eight targets as young as $\leq$ 10 Myrs. Substellar 
companions to young stars are still in early cooling phases and are more
readily detectable as a result of their elevated luminosities 
(L $\sim$ t$^{-1.3}$ M$^{2.24}$).  Among the youngest targets are  
several members of the TW Hydra Association (TWA), a loose stellar 
association of (perhaps) two dozen identified members 
(Webb \& Zuckerman 1999), all 
$\sim$ 10 Myrs of age, comprising the nearest site of recent 
star formation to the Earth (d $\sim$ 50pc). 

A Search for Low Mass Companions to Nearby M Stars. 
Twenty-seven M-dwarfs which are a) very nearby (d $\leq$ 6pc) with 
spectral types later than M3.5; b) young (age $\leq$ 10 My) with d 
$\leq$ 25pc; and c) spectrally the latest known (i.e, ``ultra-cool'' 
dwarfs later than 
$\sim$ M8.5) were observed.  For the nearest stars in this sample, 
the minimum separations for companion detections are as small as 1.2 AU. 

Dust Disks around Main Sequence Stars. Stars with large 
IRAS excesses and other indicators of the likely presence of circumstellar 
dust were observed at 1.1 and/or 1.6 microns. Shorter wavelength 
observations provide better coronagraphic sensitivity for 
circumstellar scattered light which will closely follow the spectral energy  
distributions of the central stars. While observations were tailored 
for disk detection, all images were analyzed for the possible 
presence of close low-mass companions. 

In a limited number of cases, we obtained follow-up 
observations of newly-found candidates in other
NICMOS filters before the cryogenic coolant in NICMOS was exhausted.
The color information was used to discriminate ill-placed unresolved 
background objects from likely physical companions.

\section{Extra-Solar Giant Planet Detection}
NICMOS coronagraphy enables the direct imaging of young EGPs.
This is demonstrated in the case of a candidate EGP companion 
to TWA6 (ROSAT116, H=6.9, K7V). This X-ray source is a TWA member 
and has an estimated age of $\sim$ 10 Myr, concurrent with the 
ensemble age of other association members. As shown in Figure 1 (left), 
the H=20.1 putative companion is seen with a S/N of 50 at an angular 
separation of  2\farcs5\ (125 AU projected 
distance) from its primary which is brighter by 13.2 magnitudes. In a follow-up 
NICMOS camera 1 observation at 0.9\micron\ (F090M filter) the 
object was unseen to a (less-sensitive) 3$\sigma$ limiting magnitude of $>$ 22,
 indicative of a very red source (if not a highly reddened 
background object).  If the object is physically associated and coeval 
with TWA6, its absolute H magnitude suggests an effective 
surface temperature of $\sim$  800K (for an surface gravity of 
7.5x10$^4$) and a mass of $\sim$  1 Jupiter based upon evolutionary models by 
Burrows et al (1997).  

\begin{figure}
\plotone{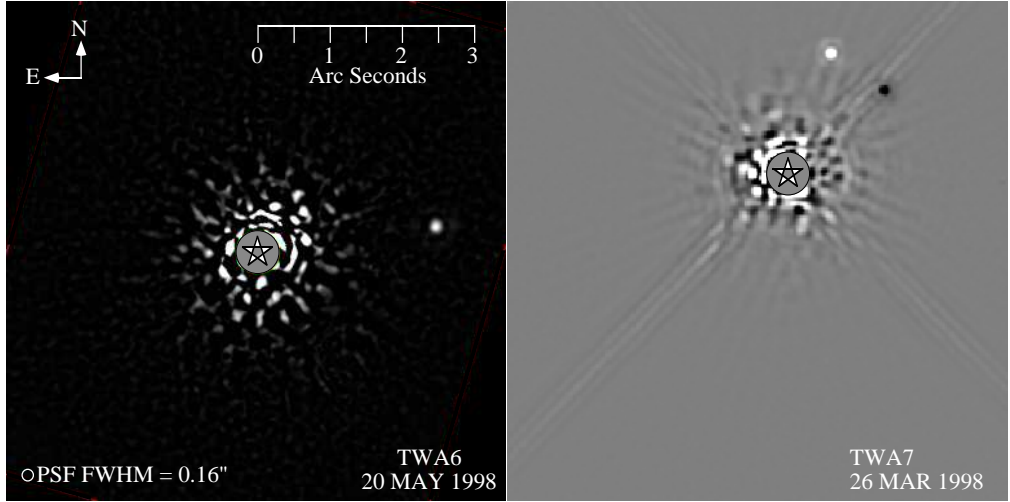}
\caption{NICMOS F160W coronagraphic images of TW Hydra association 
members TWA6 (left, reconstructed from rotated difference
image conjugates), and TWA7 (right, difference image).}
\end{figure}

In the case of TWA6``B'', as well as a 
number of other substellar companion candidates, we are 
obtaining HST/STIS 0.75-1.0 \micron\ spectra of the objects  
to help elucidate their physical natures.  We are also
attempting to measure differential proper 
motions to establish (or reject) physical association
with ground-based adaptive optics facilities. Such
astrometric measures may be difficult with the
current temporal baseline of less than two years.  Re-observations with NICMOS, when 
it is returned to service in HST Cycle 10, should unequivocally 
establish whether or not TWA ``A'' and``B'' are gravitationally bound,
and hence if TWA6``B''  is a young Jovian planet.

While the unambiguous determination of the planetary nature of a 
suspected companion demands dynamical and spectroscopic 
confirmation, the rejection of putative young EGJs may 
be done with high confidence based on their color indices.  For 
example, a more luminous candidate companion object with
an apparent H magnitude of 16.8
was identified at a similar distance from another TW Hydra 
association member, TWA7 (M1V, H=7.4) as shown in Figure 1 (right).  A 
later NICMOS observations found its  I'[F090M(0.9\micron)] - H color to 
be +0.88, too blue to be a substellar object at the 
age of the primary, and most-likely a background K star.   
One might conjecture extrinsic temporal variations in the color indices
arising, for example, from cometary impacts.  This is an unlikely scenario
as a sustained blueing of more than an order of magnitude 
would be required for a 10 Myr EGP of this H-band luminosity
to mimic a background star of the same color. 

\section{Brown Dwarfs}
Recently, many old isolated ``field'' brown dwarfs have been 
discovered from large ground-based surveys such as the 2-Micron 
All-Sky and Sloan Digital Sky surveys 
and were subsequently confirmed 
spectroscopically.  Indeed, it has been suggested that brown dwarfs, 
once considered somewhat exotic objects, may be as common as 
stars.  Yet, the companion mass fraction of these substellar objects,
which occupy a niche in the mass function between planets and 
stars, remains largely unknown.  How common are they? Do 
companion brown dwarfs form in a process more like planets than 
stars? How does the presence of a brown dwarf companion effect 
the evolution of a newly-forming solar system? Our small NICMOS 
survey begins to address these questions but many more objects will 
have to be studied before they are fully answered.

The first brown dwarf companion candidate identified from the NICMOS surveys, 
TWA5B (CD -33\deg 7796B, M1.5V, H=7.2) as reported by Lowrance et al 
(1999), is also a member of the TW Hydra association.  
The H=12.14 magnitude (Vega system) companion is readily seen in 
the NICMOS coronagraphic difference image shown in Figure 2 (left) 
with a separation of 1\farcs96 (98 AU projected distance) 
at a PA of 358\fdg9. Direct imaging with the KeckI/NIRC and
KeckII/LRIS acquisition cameras (possible given the modest primary/secondary contrast 
ratio and sufficiently large separation) show color indices of 
I-J = 3.2, I-H=3.7, and I-K = 4.4. With an absolute 
H magnitude of 8.41 (assuming physical association, so d = 50 pc) 
and a bolometric correction of BC(H) = 2.8
we ascertained its luminosity as 0.0021 L$_\odot$ and 
its effective temperature as $\sim$ 2600K.  

\begin{figure}
\plotone{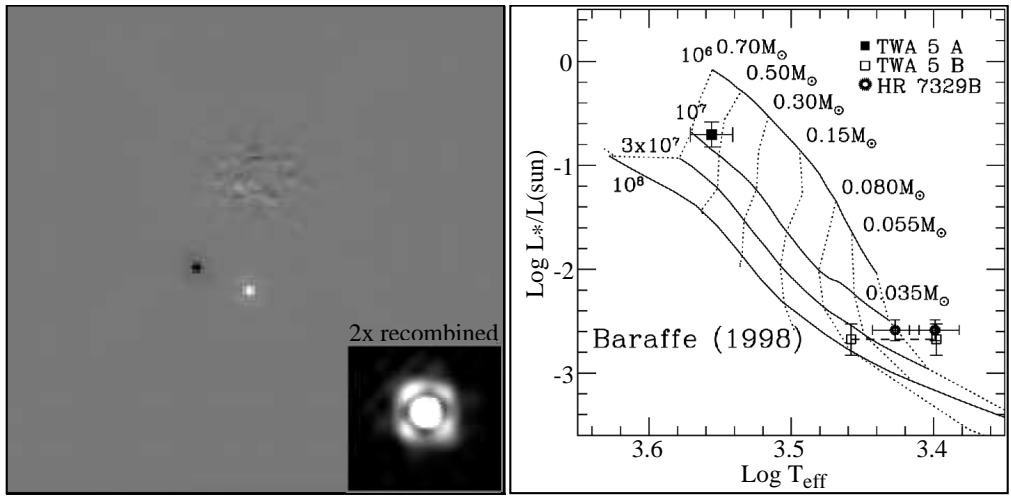}
\caption{(Left) F160W coronagraphic difference image of TWA5B 
(recombined image inset). (Right) Placement of TWA5A/B \&
HR7329B on evolutionary tracks.}
\end{figure}

Young, and therefore hot, brown dwarfs have spectra resembling M 
dwarf stars of the same photospheric temperatures.  Thus we 
photometrically classify TWA5B as M8.5V. 
Using evolutionary tracks from Baraffe (1998) (see Figure 
2, right) and others we estimate the mass of the object 
to be $\sim$  0.02 M$_\odot$
(20 Jupiters).  We recently obtained a short-wavelength NIR 
spectrum using the G750M grating in STIS.  We compare the 
8000-9000\AA\ spectrum of TWA5B 
to that of two late M-dwarf standards, J2309+1594 (M8.5V) and 
J1239+2029 (M9V), (see Figure 3) for which we find a very good fit 
confirming our photometrically determined photospheric temperature and spectral classification.


\begin{figure}[h]
\includegraphics{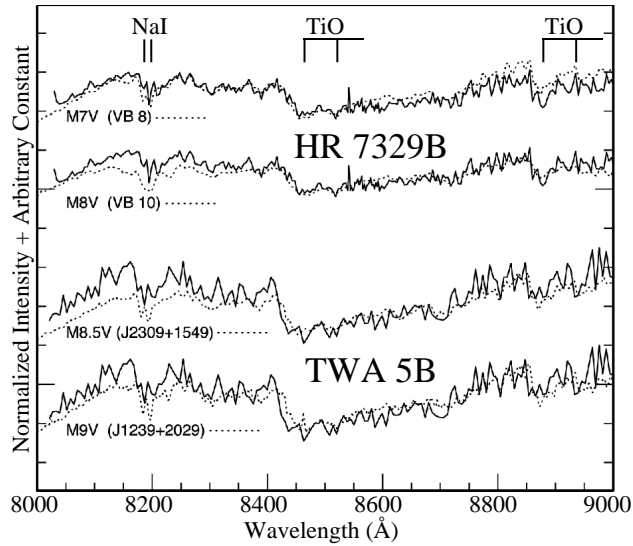}
\vspace{7.0truecm}
\caption{STIS spectra of TWA5B, HR7329B, and late M-dwarfs}
\end{figure}


We compare TWA5B with a second brown dwarf companion 
identified with NICMOS and confirmed spectroscopically with STIS, 
HR 7329B (Lowrance, et al 2000).  The older 
($\sim$30 Myr) A0V star (H=5.05) is 
at a comparable distance of 47pc.  Assuming companionship for the 
4\farcs17\ distant unresolved object (196 AU projected distance), 
the H=11.90 companion candidate imaged by NICMOS at a PA of 
166\fdg8 was estimated to have a luminosity of 0.0026 L$_\odot$ 
(Habs = 8.54 with BC(H)=2.67). As with CD -33\deg 7795 
we place HR 7329B on evolutionary models and find a mass of 
$\sim$ 0.04M$_\odot$ (40 Jupiters). Its spectrum, from 
which we deduce a spectral class of M7.5V, is shown in Figure 3 
along with TWA5B and the M7V and M8V spectral 
standards VB8 and VB10, respectively.

\ 

In conclusion, we look forward to the return of NICMOS in HST cycle 10 which will permit an expansion of this stellar survey, and unambiguous astrometric follow-up of our already identified companion candidates.

\acknowledgements
This work is based on observations with the NASA/ESA Hubble Space
Telescope, obtained at STScI, which
is operated by AURA, Inc., under NASA contract NAS2-6555 and
supported by NASA grants NAG5-3042 and GO-98.8176A to the
NICMOS IDT and EONS teams.


\begin{references}
\reference Baraffe, I., Chabrier, G., Allard, F., \& Hauschildt, P. H. 
1998 A\&A, 337, 403

\reference Burrows, A., Marley, M., Hubbard, W. B.,
Lunine,J. I., Guillot, T., Saumon, D., Freedman, R., Sudarsky, D.,
Sharp, C. 1997 ApJ, 491, 856


\reference Lowrance, P.J., Becklin, E.E., Schneider, G., Hines, D., 
Kirkpatrick, D., Koerner, D., Low, F., McCarthey, D., Meier, R., Reike, 
M., Smith, B.A., Terrile, R., Thompson, R., Zuckerman, B. 1998, Conf Proc., 
NICMOS and the VLT, ed. W. Freufling and R. Hook (Garching: ESO), 96

\reference Lowrance, P., Becklin, E., Schneider, G., Kirkpatrick, D., 
Zuckerman, B., Plait, P., Malimuth, E., Heap, S., Weinberger, A., 
Smith, B., Terrile, B., Schultz, A., and Hines,
D., 2000, ApJ, in press

\reference Lowrance P.J., McCarthy, C., Becklin, E., Zuckerman, B., 
Schneider, G., Webb., R., Hines, D., Kirkpatrick, J., Koerner, D., Low, F., 
Meier, R., Rieke, M., Smith, B., Terrile, R., \& Thompson, R. 1999, ApJ, 512 , L69 

\reference Schneider, G. 1998 Conf Proc., 
NICMOS and the VLT, ed. W. Freufling and R. Hook (Garching: ESO), 88

\reference Schneider, G., Thompson, R.I., Smith, B.A.,\& Terrile, R.J, 
1998, SPIE Conf. Ser. Vol 3356, Space Telescopes and Instrumentation V, 
ed. P. Bely \& J. Breckenridge (Bellingham:SPIE) 222

\reference Webb, R.A. \& Zuckerman, E., 1999 BAAS, 195, 3202W

\reference Weinberger, A.J., 2000, in ASP Conf. Ser. Vol. TBD, 
Disks, Planetesimals and Planets, ed. F. Garz—n, Carlos Eiroa, 
Dolf de Winter and T. J. Mahoney (San Francisco: ASP), these proceedings.
\end{references}
\end{document}